\documentstyle[aps,twocolumn]{revtex}

\begin{document}
\title{Hole-Hole Interaction Effect in the Conductance of the Two-Dimensional Hole
Gas in the Ballistic Regime}
\author{Y.Y. Proskuryakov$^1$, A.K. Savchenko$^1$, S.S. Safonov$^1$, M. Pepper$^2$,
M.Y. Simmons$^2$, D.A. Ritchie$^2$}
\address{$^1$ School of Physics, University of Exeter, Stocker Road,\\
Exeter, EX4 4QL, U.K.\\
$^2$ Cavendish laboratory, University of Cambridge, Madingley Road,\\
Cambridge CB3 0HE, U.K.}
\maketitle

\begin{abstract}
On a high mobility two-dimensional hole gas (2DHG) in a
GaAs/GaAlAs heterostructure we study the interaction correction to
the Drude conductivity in the ballistic regime, $k_BT\tau /\hbar $
$>1$. It is shown that the 'metallic' behaviour of the resistivity
($d\rho /dT>0$) of the low-density 2DHG is caused by hole-hole
interaction effect in this regime. We find that the temperature
dependence of the conductivity and the parallel-field
magnetoresistance are in agreement with this description, and
determine the Fermi-liquid interaction constant $F_0^\sigma $
which controls the sign of $d\rho /dT$.
\end{abstract}


It is well known that electron-electron interaction gives rise to a quantum
correction to the classical (Drude) conductivity caused by impurity
scattering \cite{AltshulerB}. Its manifestation can be quite different in
the two regimes which relate the quasi-particle interaction time, $\hbar
/k_BT$, to momentum relaxation time, $\tau $: diffusive ($k_BT\tau /\hbar <1
$), and ballistic ($k_BT\tau /\hbar $ $>1$). So far the interaction
correction to the conductivity of two-dimensional (2D) systems has been
studied, both theoretically and experimentally, only in the diffusive
regime, which is applicable to low-mobility (small $\tau $) systems \cite
{AltshulerB}. Experimentally, the interaction correction was seen to be
negative and produce a logarithmic decrease of the resistivity with
increasing temperature, similar to the interference correction due to weak
localisation (WL). Theory, however, suggests that the sign of the
interaction effect in the diffusive regime can be different, dependent on
the value of the interaction constant $F_0^\sigma $. Thus the correction can
become positive and give rise to a 'metallic' temperature dependence with $%
d\rho /dT>0$ \cite{Finkelstein}.

The role of interactions in the conductance of 2D systems has now been
intensely discussed, after the observation of the'metallic' behaviour in
some low-density, high-mobility 2D systems \cite{many}. If one attempts to
apply the conventional interaction theory, it can be seen that in high
mobility structures the diffusion approximation becomes invalid even at low $%
T$. Recently, a theory of the interaction correction in the {\em ballistic}
and intermediate regimes has been developed \cite{Aleiner}. Stimulated by
this theory, in this work we examine the role of the hole-hole interaction
effects in a 2D hole gas (2DHG) which shows a 'metallic' $\rho (T)$. We
analyse the temperature dependence of the conductivity and positive
magnetoresistance in parallel field, and show that these two main features
of the 'metallic' state can be explained by the interaction effect. It has
the same origin as the logarithmic correction studied earlier \cite
{AltshulerB}, but now manifests itself in the ballistic regime.

The interaction theory \cite{Aleiner} considers elastic (coherent) electron
scattering on the modulated density of other electrons (Friedel oscillation)
caused by an impurity with a short-range potential. The phase of the Friedel
oscillation, $\Delta p\propto \exp (i2k_Fr)$, is such that the wave
scattered from the impurity interferes constructively with the wave
scattered from the oscillation, Fig. 1 (a), leading to the quantum
correction to the Drude conductivity $\sigma _0$. The model \cite{Aleiner}
gives several predictions to be tested experimentally.

{\em Firstly}, the logarithmic correction in the diffusive regime of
multiple impurity scattering, becomes a linear temperature dependence in the
case of a single scatterer, at $k_BT\tau /\hbar >1$:

\begin{eqnarray}
\delta \sigma \left( T\right) &=&\frac{e^2}{\pi \hbar }\frac{k_BT\tau }\hbar
\left( 1+\frac{3F_{\text{0}}^\sigma }{1+F_{\text{0}}^\sigma }\right)
\label{eq1} \\
\ &=&\sigma _0\left( 1+\frac{3F_{\text{0}}^\sigma }{1+F_{\text{0}}^\sigma }%
\right) \frac{k_BT}{E_F}  \nonumber
\end{eqnarray}
where $F_0^\sigma $ is the Fermi liquid interaction parameter in the triplet
channel. The coefficient in the temperature dependence originates from two
contributions: one due to exchange processes (Fock) and another due to
direct interaction (Hartree). Similar to the diffusion regime, the sign of $%
d\rho /dT$ depends on the constant $F_{\text{0}}^\sigma $. It is important
that for a given $F_{\text{0}}^\sigma $, dependence $\rho \left( T\right) $
can be 'metallic' even when the logarithmic correction at smaller $k_BT\tau
/\hbar $ is 'insulating'. It is also interesting to note that according to
\cite{Aleiner} the actual transition to the ballistic regime occurs at $%
0.1k_BT\tau /\hbar $, so that experiments on high-mobility structures can be
easily driven into the ballistic regime.

{\em Secondly}, for a wide range of parameter $F_{\text{0}}^\sigma $ the
model allows the change of the sign of $d\rho /dT$ with parallel magnetic
field - the effect seen in recent experiments. Magnetic field suppresses the
correction in the triplet channel in Eq. (\ref{eq1}), resulting in a
universal, positive correction to the Drude conductivity in magnetic field, $%
\sigma _0^B$, and hence the 'insulating' behaviour of $\rho (T)$:

\begin{equation}
\delta \sigma =\sigma _0^B\frac T{T_F}\text{ at }B\geq B_S.  \label{eq2}
\end{equation}
Here $B_S$ is the field corresponding to the full spin polarisation of the
2D system, $B_S=2E_F/g^{*}\mu _B$, where $g^{*}$ is the Lande g-factor, $\mu
_{B\text{ }}$ is the Bohr magneton, and $T_F$ is the Fermi temperature. Note
that the same functional dependence as in Eq. (\ref{eq1}) was derived in
\cite{Gold}, where the authors consider the same physical phenomenon in
terms of the temperature effect on screening of the impurity potential. The
model \cite{Gold} has been applied to the analysis of the linear $\rho (T)$
in several experiments on Si MOSFETs and GaAs structures, although no
quantitative agreement with the model was achieved. It is important to
mention that model \cite{Gold} considers only the Hartree potential of
interacting electrons and ignores the Fock contribution. As a result, it
only predicts the positive sign of $d\rho /dT$ and does not allow the change
in $d\rho /dT$ with magnetic field.

The Fermi-liquid constant $F_0^\sigma $ has a significant physical
meaning. It can be considered as the ratio of exchange and kinetic
energies, and it also comes into the magnetic susceptibility
\cite{Landau}:

\begin{equation}
\chi (n)=\frac{\chi _0}{1+F_0^\sigma }.  \label{eq3}
\end{equation}
Recently, there have been reports that in the 'metallic' state of the 2DEG
in Si-MOSFETs the g-factor diverges when approaching the
'metal'-to-'insulator' transition \cite{Shashkin}, indicating the
ferromagnetic (Stoner) instability expected in low-density 2D systems \cite
{Ando}. In our 2DHG the parameter $r_s$ is twice as large as in \cite
{Shashkin}, $r_s\sim 20$ near the crossover, so that we can also expect a
manifestation of the Stoner instability. Therefore, our analysis of the
conductance in terms of Eq. (\ref{eq1}) can give the value of the
interaction parameter $F_0^\sigma $ and show how close to the Stoner
instability the system is - from Eq. (\ref{eq3}) the instability is expected
to occur at $F_0^\sigma =-1$.

The experiments have been performed on a (311)A GaAs/AlGaAs heterostructure
with a peak mobility of $6.5\times 10^5$ cm$^{\text{2}}$V$^{\text{-1}}$s$^{%
\text{-1}}$, which shows the crossover from 'metal' to 'insulator' at $p\sim
1.5\times 10^{10}$cm$^{-2}$, Fig. 1 (b). A standard four-terminal lock-in
technique has been used for resistivity measurements at temperatures down to
50 mK, with currents of 1-10 nA to avoid electron heating. The hole density $%
p$ in the 'metallic' region is varied by the gate voltage in the range $%
(2.09-9.4)\times 10^{10}$cm$^{-2}$ , corresponding to the interaction
parameter $r_s=10-17$ (with the effective mass $m^{*}$ taken as $0.38$$m_e$
\cite{Stormer}). In \cite{Sivan,Murzin} the 'metallic' character of a higher
density 2DHG in GaAs was explained in terms of inelastic scattering between
two hole subbands, split due to spin-orbit interaction. In our low-density
structures the effect of the band splitting is negligible.

Fig. 1 (b) represents the temperature dependence of the resistivity, with
the 'metallic' region under study marked by a box. The increase of the
resistivity with $T$ can be simply due to phonon scattering, which cannot be
ignored in GaAs structures with piezo-electric coupling even at temperatures
below 1K. In Fig. 1 (c,d), curves $\rho (T)$ for different densities are
plotted together with the theoretical dependence presented as $\rho (T)=\rho
_0+\rho _{ph}$, where $\rho _0=\sigma _0^{-1}=\rho \left( T=0\right) $ is
the residual resistivity due to impurity scattering, obtained by
extrapolation to $T=0$, and $\rho _{ph}$ is the result of the calculations
for the phonon scattering in GaAs heterostructures \cite{Karpus}. The latter
is represented as $\rho _{ph}(T)=\frac{a(T/T_0)^3}{1+c(T/T_0)^2}$, where
parameters $a$ and $c$ depend on the carrier density, effective mass and
crystal properties, and $T_0=k_B^{-1}\sqrt{2m^{*}S_t^2E_F}$, where $S_t$ is
the transverse sound velocity. This relation corresponds to the intermediate
temperature range between the Bloch-Gruneisen, $T<T_0$, and the linear, $%
T>T_0$, regimes, for the case of non-screened phonon scattering. (The
criterion $T<T_0/\pi $ \cite{Karpus2} for the screened phonon scattering is
not satisfied for the majority of our data.) One can see that at the highest
$p$, phonon scattering can fully explain the experimental dependence $\rho
(T)$. However, with decreasing density another contribution develops, which
dominates at low $T$ and low densities. Fig. 2 (a) shows this contribution
obtained by subtracting that of phonon scattering. The peak-like shape of $%
\rho (T)$, with the maximum at $T_{\max }\approx 0.3T_F$, is in qualitative
agreement with numerical calculations in \cite{DasSarmaHwa} where it is
explained by the transition to the non-degenerate regime at $T>0.3T_F$.
(Similar $\rho (T)$ dependence with a peak has also been seen in \cite{Mills}%
).

In order to compare the results in the low-temperature range of $\rho (T)$
with Eq. (\ref{eq1}), we replot in Fig. 2 (b) the data in conductivity form
: $\Delta \sigma (T)=(\rho (T)-\rho _{ph}(T))^{-1}-\rho _0^{-1}$. The
condition for the ballistic regime $k_BT\tau /\hbar \geq 1$ is satisfied in
our structure at $T>50-100$ mK, and a linear fit of $\Delta \sigma (T)$
gives the value of parameter $F_0^\sigma $, which is presented in Fig. 2c
for different $p$. The following comments can be made on this result. {\em %
Firstly}, the interaction constant is negative and this provides the
'metallic' slope in $\rho \left( T\right) $.{\em \ Secondly}, its absolute
value decreases with increasing density, which is in agreement with the
expectation that the ratio of the exchange to kinetic energy of
quasi-particles decreases to zero at large densities. {\em Thirdly}, one can
see that the measured value does not exceed 0.42, and when extrapolated to
the density of the crossover from 'metal' to 'insulator' ($p\sim 1.5\times 10
$ cm$^{-2}$), is much smaller than the value of $\left| F_o^\sigma \right| =1
$ expected for the Stoner instability. This implies that our description of
the 'metallic' system as a weakly interacting Fermi-liquid is
self-consistent.

Let us now turn to the increase of resistance with parallel field shown in
Fig. 3 (a), which is similar to that observed earlier on the 2DHG \cite
{Yoon-Papadakis}. There is a characteristic feature in the data - a bend,
which shifts towards smaller fields as the density is decreased. It was
recently shown that this hump corresponds to the magnetic field $B_S$ of
full spin polarisation of the 2DHG \cite{Tutuk}. However, there has been no
quantitative analysis of the magnetoresistance in the 2DHG, which we now
attempt. We start with the analysis which is similar to that performed on 2D
electrons in a Si MOSFET \cite{Shashkin} and GaAs heterostructure \cite
{Dolgopolov-Harpai}. It is based on the model \cite{Dolgopolov-Gold} of
positive magnetoresistance at $T=0$, which considers the effect of parallel
field on the screening of the impurity potential, affected by the presence
of two spin subbands with different Fermi momenta $k_F$. The fact that this
model does not consider the Fock component in the interacting potential
should not now affect the magnetoresistance result, as the magnetic field
only acts on the Hartree term. Fig. 3 (b) shows $\rho (B_{||})/\rho
(B_{\Vert }=0)$ as a function of dimensionless magnetic field $B/B_S$, with $%
B_S$ found as a fitting parameter. Its value is shown by the dashed line in
Fig. 3 (a) and corresponds to the position of the hump. In accordance with
\cite{Dolgopolov-Gold}, all the data in the density range $p=1.43-8.34\times
10^{10}$cm$^{\text{-2}}$ collapse on one curve. There is satisfactory
agreement with the experiment, apart from the region close to $B_S$ where we
can expect a contribution of another mechanism taking place, which takes
into account a finite thickness of the conducting layer \cite{DasSarma-parB}.

Using the value of $B_S$ one can obtain the effective $g$-factor, using the
relation $g^{*}=2E_F/\mu _BB_S$, whose dependence on the density is shown in
the inset to Fig 3 (b). Note that contrary to \cite{Shashkin}, where a
similar analysis on a 2DEG in Si structures showed a rapid increase of the $g
$-factor with decreasing density near the 'metal'-to-'insulator' transition,
the $g$-factor in our case decreases with decreasing density. Similar
behaviour was recently observed for 2D electrons in GaAs \cite{Tutuc2}.

Fig. 4 (inset) shows the temperature dependence of the magnetoresistance,
where one can see that $B_{||}$ drives the 'metallic' state into
'insulator'. To compare this result with the prediction given by Eq. (\ref
{eq2}) we analyse the temperature dependence of the resistivity at field $B_S
$. (In this analysis we neglected altogether the phonon contribution, as at $%
T<0.5$ K it becomes less than 5\%, due to the four-fold resistance increase
at $B_{||}=B_S$). The resulting dependences, Fig. 4 (a), are indeed linear,
in accordance with Eq. (\ref{eq2}). By extrapolation to $T=0$, we find the
value of the Drude conductivity $\sigma _0^B$  and determine the slope $%
\alpha $ of the straight lines. Its value is close to the expected one, $%
\alpha $=1, at all studied densities, although we find that agreement is
better for smaller $p$, where $\alpha $ increases to 0.92. Such a behaviour
of $\alpha $ can be attributed to the fact that in the real system the
scatterers are not point-like, as assumed in the theory. However, with
decreasing density and increasing Fermi wavelength, $\lambda _F\propto
p^{-1/2}$, the approximation of short-range scatterers becomes more
applicable.

The discussed model also gives a simple prediction for the
magnetoconductivity in the ballistic regime $\Delta \sigma =\sigma
(B_{||},T)-\sigma (0,T)$ in weak fields, such that $x=\frac{Ez}{2k_BT}\leq
1+F_0^\sigma $, which can also be tested experimentally \cite{AleinerB}:

\begin{equation}
\Delta \sigma (B_{||})=\frac{2F_0^\sigma }{1+F_0^\sigma }\sigma _0\frac T{T_F%
}K_b\left( \frac{Ez}{2T},F_0^\sigma \right) ,  \label{eq4}
\end{equation}
where $Ez=g^{*}\mu _BB_{||}$ and $K_b\left( x,F_0^\sigma \right) \approx
x^2f(F_0^\sigma )/3$, $f(z)=1-\frac z{1+z}\left[ \frac 12+\frac 1{1+2z}-%
\frac 2{(1+2z)^2}+\frac{2\ln (2(1+z))}{(1+2z)^3}\right] $.

In Fig. 4 (b) we plot the magneto-conductivity at $T=0.6$ K as a function of
$B_{||}^2$ for the fields satisfying the above condition. We use $\sigma _0$
obtained in the above analysis at $B_{||}=0$, and $g^{*}$ determined from
the analysis of $\rho (B_{||})$ at the lowest $T$. Therefore, the only
unknown parameter in the slope of $\Delta \sigma (B_{||}^2)$ is $F_0^\sigma $%
. We extract its value for different densities and compare it with that
determined earlier from $\rho (T)$ at zero field, Fig. 2 (c). Good agreement
between the results of the two approaches proves the validity of the
interpretation of the 'metallic' state.

In conclusion, we have studied the temperature dependence of the
conductivity and the magnetoresistance in parallel field of a low-density
(large $r_s$) 2D hole gas in the 'metallic' phase, near the crossover in the
sign of $\rho \left( T\right) $. We have demonstrated that the 'metallic'
character of $\rho \left( T\right) $ and the positive magnetoresistance are
caused by the hole-hole interaction in the ballistic limit $k_BT\tau /\hbar
>1$. We have found the Fermi liquid constant $F_0^\sigma $, which determines
the sign of $\rho (T)$. Its value near the crossover appears to be
significantly smaller than expected for the ferromagnetic instability.

We are grateful to I.L. Aleiner, B.L. Altshuler and B.N. Narozhny for
stimulating discussions, and EPSRC and ORS award funds for financial support.


\begin{references}
\bibitem{AleinerB}  G. Zala, B. N. Narozhny, and I. L. Aleiner (in
preparation). \\

Figure captions:\\

Fig.1. (a) Diagram of electron scattering by an impurity and the Friedel
oscillation it produced.

(b) Temperature dependence of the resistivity at different hole densities
near the crossover in the sign of $d\rho /dT$.

(c,d) Resistivity on the 'metallic' side of the crossover (symbols), and
contribution to $\rho (T)$ due to phonon scattering (solid lines).\\

Fig. 2 (a) Impurity scattering contribution $\Delta \rho $ against
dimensionless temperature at different $p$. (For clarity, curves
in (a) and (b) are offset vertically from zero value at $T=0$.)
(b) The same data as in (a) but in conductivity form, with linear
fitting. (c) Fermi liquid parameter versus hole density. Open
symbols show the result obtained from the analysis of $ \rho (T)$
at zero magnetic field; closed symbols show the result from the
analysis of the parallel-field magnetoresistance, Fig. 4 (b).\\

Fig. 3 (a) Dependence of the resistivity on parallel magnetic field at $T=50$
mK and $p=$1.43; 1.57; 1.75; 2.03; 2.26; 2.49; 2.83; $3.36\times 10^{\text{10%
}}$ cm$^{\text{-2}}$, from bottom to top.

(b) Scaled data, with an added curve $\rho (B_{||})$ for $%
p=8.34\times $10$^{\text{10}}$ cm$^{\text{-2}}$; solid line is the
result of the model \cite{Dolgopolov-Gold}. Inset: dependence of
the effective $g$-factor on the
hole density, obtained from the value of $B_S$.\\

Fig. 4. (a) Temperature dependence of the conductivity at
$B_{||}=B_S$ for different hole densities. Coefficient $\alpha
=0.53$ is obtained for $p=2.49$, $2.83\times 10^{\text{10}}$
cm$^{\text{-2}}$ ; $\alpha =0.62$ for $p=2.26\times 10^{
\text{10}}$ cm$^{\text{-2}}$ ; $\alpha =0.74$ for $p=2.03\times
10^{\text{10} }$ cm$^{\text{-2}}$; and $\alpha =0.92$ for
$p=1.43$, $1.57$, $1.75\times 10^{\text{10}}$ cm$^{\text{-2}}$.
Inset: $\rho (B_{||})$ for $p=2.26\times $ 10$^{\text{10}}$
cm$^{\text{-2}}$, at different temperatures: $T=0.1$, $0.2$,
$0.3$, $0.45$, $0.6$, $0.8$ K.

(b) Magnetoconductivity against $B_{||}^2$, at $T=0.6$ K for
densities $p=2.03$, $2.26$ , $2.49$, $2.83$, $3.36\times
10^{\text{10}}$ cm$^{\text{-2}}$.
\end{references}
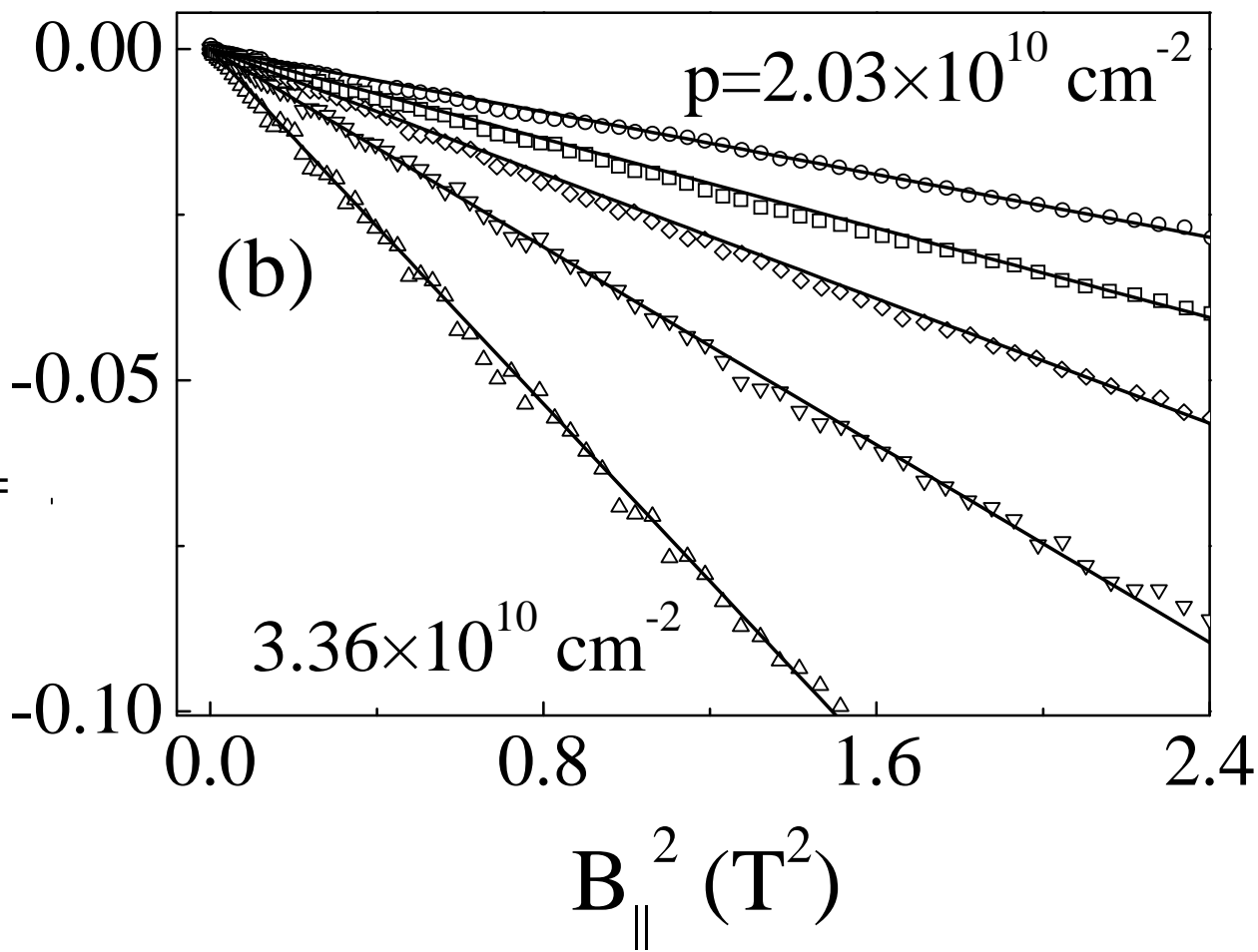
\end{document}